\pdfoutput=1
\documentclass[aps,prb,twocolumn,10pt]{revtex4-1}
\usepackage[utf8]{inputenc}
\usepackage{libertine}
\usepackage[libertine]{newtxmath}
\usepackage{mathtools,graphicx,microtype,bm}
\usepackage[cal=boondoxo,frak=boondox]{mathalfa}
\usepackage[colorlinks=true,allcolors=blue]{hyperref}

\begin{document}
\title{Radiative Decay of Bound Electron Pairs into Unbound Interacting  Electrons in 2D Materials with Two-Band Spectrum}

\author{Bagun S.\ Shchamkhalova and Vladimir A.\ Sablikov}
\affiliation{Fryazino Branch of the Kotelnikov Institute of Radio Engineering and Electronics, Russian Academy of Sciences, Fryazino, Moscow District, 141190, Russia}

\begin{abstract}
Bound electron pairs (BEPs) arising due to peculiarities of the band structure of topologically non-trivial materials  are of interest as charge and spin carriers with energies in the band gap. Moreover, being composite bosons, they can also possess completely unusual collective properties. In this regard, of highest importance is the problem of their decay time. The processes of radiative decay of BEPs are studied, considering the electron–electron interaction in the final state, into which the the BEPs decay. It is found that the decay time of singlet BEPs lies in the nanosecond range and significantly exceeds other characteristic electron relaxation times. The radiative decay of triplet BEPs is impossible.
\end{abstract}

\maketitle

\section{Introduction}

Bound electron pairs (BEPs) appearing despite their Coulomb repulsion have attracted a great deal of interest for a long time mainly in connection with problems of  superconductivity~\cite{combescot2015excitons, kagan2013modern, RevModPhys.62.113}. But recently, a new direction of these studies has arisen related to a possibility of Coulomb pairing of electrons in such modern materials as topological insulators (TIs)~\cite{Sablikov}, Dirac semimetals~\cite{Portnoi}, graphene and bigraphene~\cite{Sabio2010, Lee2012, MahmoodianEntinEPL2013, MarnhamShytovPRB2015}, carbon nanotubes~\cite{Hartmann}, two-dimensional systems with strong spin-orbit interaction~\cite{PhysRevB.98.115137}. Similar bound states of two fermions are also studied in systems of cold atoms in optical traps, where they are called doublons~\cite{winkler2006repulsively,PhysRevB.89.195119,Han_2016}. 

An electron pairing in most systems is caused by the peculiarities of the electronic spectrum, which results in a negative effective reduced mass of two particles. An idea of such a pairing dates back to the work of Gross, Perel', Shekhmamet'ev~\cite{Gross}, in which BEPs formed by quasiparticles with an energy near the maximum of the band spectrum were observed. Further studies have shown that in materials with a more complicated structure of electronic states, the  probability of a negative reduced mass to appear is much higher. In these materials, the electronic states are characterized not only by a spin, but also by atomic orbitals (pseudospin). The minimal, four-band model contains two such orbitals. The presence of orbital degree of freedom provides many opportunities for the formation of a negative effective reduced mass. Therefore, many types of BEP states can exist in TIs and graphene.

BEP is a new and as yet poorly studied type of composite quasiparticles, which can have quite unexpected properties. BEPs are bosons with a charge 2$e$ and a spin 1 or zero with energies in a bandgap. Therefore, BEPs can transfer the charge and spin when band electrons do not. Unusual collective phenomena can also be expected in a system with BEPs, since they are charged composite bosons. Thus, the  study of transport properties and collective phenomena arising due to BEPs is a challenge. The first step to be taken in this direction is to find out how long BEP can live. 

In a recent paper~\cite{Sablikov1}, we developed a model of the BEP radiative decay, which allowed us to estimate their intrinsic radiative lifetime. But an important question of how the repulsive interaction of electrons in the final state into which the BEP decays, affects the probability of transition  remains unclear. In the problem of BEP decay, this interaction plays an important role, since the spatial correlation of electrons in the final state largely determines the overlap of the two-particle wave functions of the initial and final states.

This question is specific for two-electron complexes, but does not arise for excitons, since both quasiparticles disappear upon decay of exciton. This is a feature due to which the decay of BEP is  very different from the decay of excitons, well studied in the literature~\cite{Hanamura, Andreani, Robert, Glazov}. It can be  expected, that if the electron-electron (e-e) interaction in the final state is  repulsive, then the decay probability decreases, but if the interaction is  attractive, the decay probability increases with increasing interaction strength. The answer to the question of whether electrons attract or repel each other is not trivial, because the relative motion of the particles in the Coulomb field is determined by the reduced effective mass, the sign of which depends on the orbital composition of the wave function of the pair unbound electrons. The latter in its turn is determined by solving the corresponding equation of motion.

In this paper, we propose a detailed theory of radiative decay of BEPs in two-dimensional (2D) materials described by the four-band model of Bernevig, Hughes, Zhang (BHZ)~\cite{BHZ}. This is a rather general model which is applied to many materials in both topological and trivial phases. We develop an approach to study the effect of the e-e interaction on a pair of unbound electrons and the BEPs decay rate. We come to two fairly general conclusions. First, pairs of bound electrons decay much more slowly than excitons in direct band semiconductors. This happens precisely because the wave function of the final state is strongly delocalized. Second, the interaction of electrons in an unbound state, into which the pair transits after decay, leads to an ambiguous effect: it can both increase and reduce the probability of the BEP decay.

The structure of the paper is as follows. In Sec.I we describe briefly both bound and unbound two-electrons states involved in BEPs radiative decay. Sec.II presents the study of intrinsic radiative decay of BEPs. The Appendix gives the details of the model and calculations.

\section{Bound and unbound pairs of interacting electrons}

The two-electron problem is considered here for materials described by the symmetric BHZ model~\cite{BHZ}. The model presents single-particle electronic states in the frame of the $\mathbf{k\cdot p}$ theory using four-band basis $\phi=(|E\uparrow\rangle,|H\uparrow\rangle,|E\downarrow\rangle,|H\downarrow\rangle)^T$, where $|E\uparrow\rangle$ and $|E\downarrow\rangle$ are the superpositions of the electron- and light-hole states with the moment projection $m_J=\pm 1/2$; $|H\uparrow\rangle$ and $|H\downarrow\rangle$ are the heavy-hole states with $m_J=\pm 3/2$. The single-particle Hamiltonian that determines the spinor of the envelope functions reads
\begin{equation}
 \hat{H}_0(\mathbf{\hat{k}})=
\begin{pmatrix}
 \hat{h}(\mathbf{\hat{k}}) & 0\\
 0 & \hat{h}^*(-\mathbf{\hat{k}})
\end{pmatrix}
\label{H_0}
\end{equation}
\begin{equation*}
\hat{h}(\mathbf{\hat{k}})=
\begin{pmatrix}
 M\!-\!B\hat{k}^2 & A\hat{k}_+\\
 A\hat{k}_- & -M\!+\!B\hat{k}^2
\end{pmatrix}\,,
\end{equation*}
where $\mathbf{\hat{k}}$ is the quasi-momentum operator, $\hat{k}_+=\!\hat{k}_x\!+\!i\hat{k}_y$, $\hat{k}_-= \hat{k}_x\!-\!i\hat{k}_y$.  $A$, $B$, and $M$ are the BHZ model parameters: $A$ describes the hybridization of the electron and hole bands, $M$ is the mass term, $B$ is the parameter of the dispersion in the electron and hole bands, which are assumed to be symmetric. In what follows we use dimensionless units by normalizing all energies to $|M|$ and distances to $\mathbf{r}_i = \sqrt{|B/M|}$. In the topological phase $MB > 0$ and the parameter $a= A/\sqrt{|MB|}$ essentially affects the dispersion in the conduction and valence bands: for $a > 2$, the band dispersion is quadratic near the extremes with positive effective mass at the conduction band bottom, for $a =\sqrt{2}$  the band dispersion is nearly flat at the extremes of the bands and for $a <\sqrt{2}$ the dispersion has a Mexican-hat shape.
 
The Hamiltonian of two interacting electrons has the form
\begin{equation}
 \hat{H}(1,2)=\hat{H}_0(\mathbf{\hat{k}}_1)\oplus\hat{H}_0(\mathbf{\hat{k}}_2)+V(\mathbf{r}_1-\mathbf{r}_2)\cdot\hat{\mathbf{I}}_{16\times 16}\,,
\label{H_2}
\end{equation}
where $V(\mathbf{r})$ is a pair e-e interaction potential.

Two-particle wave functions are represented by a 16-rank spinor that defines the envelope functions in the basis $\phi \,\otimes \,\phi$. Within the BHZ model the relative motion and the motion of the center-of-mass are not separable. Nevertheless, since the system is translational invariant, in the center-of-mass frame with $\mathbf{R}=(\mathbf{r}_1+\mathbf{r}_2)/2$ and $\mathbf{r}=\mathbf{r}_1-\mathbf{r}_2$ the wave function can be written in the form:
\begin{equation}
 \Psi(\mathbf{R},\mathbf{r})=\sum_m\Psi_{m,\mathbf{K}}(r,\varphi)e^{im\varphi}\cdot e^{i\mathbf{KR}},\,
\end{equation}
where $\mathbf{K}$ is the total momentum of the pair and the functions $\Psi_{m,\mathbf{K}}(r,\varphi)$ describe the relative motion, with $r$ and $\varphi$ being polar coordinates relative to the center of mass. For a given quantum number $m$ there can be several functions corresponding different radial quantum numbers.

In this article, we will focus on the case when in the bound state $ \mathbf {K} = 0 $ and the total momentum of the unbound electrons is small, $ \mathbf {K} \ll 1 $. This is justified because the momentum of unbound pairs of electrons is equal to the momentum of a photon, which is very small. The calculation of bound states is performed in the same way as in Ref.~\onlinecite{Sablikov}. When calculating unbound states, we consider the $ K $ -dependent part of the Hamiltonian~(\ref{H_2}) as a perturbation.

As the Hamiltonian~(\ref{H_2}) does not contain any spin-dependent terms, the $z$-component of the spin $S_z$ is conserved. Therefore, there are two kinds of eigenstates with different spins, which are singlets with $S_z=0$ and triplets with $S_z=\pm 1$. Because of the spin conservation, the Schr\"odinger equation with Hamiltonian~(\ref{H_2}) splits into uncoupled systems of equations for four-rank spinors of the singlet and triplet states~\cite{Sablikov}.   

Here we consider in detail the singlet bound and unbound states described by a spinor of the form ${\Psi}_s(\mathbf{r})=\left(0,0,\psi_3,\psi_4,0,0,\psi_7,\psi_8\right)^T$ in both trivial and topological phases. The calculation of the wave functions $\Psi_s$ is described in Appendix~\ref{App_wave_func}. Triplet wave functions are calculated in a similar way, but the details of their calculation are omitted, since these states, as we have found, do not interact with photons and therefore are dark.

The interaction potential is modeled by a step function:
\begin{equation}
v(r)= v_0 \Theta (r_0-r)\,,
\end{equation}
where $v_0$ and $r_0$ are the amplitude and radius of the e-e interaction potential.

\begin{figure}[htb]
\centerline{\includegraphics*[width=1\linewidth]{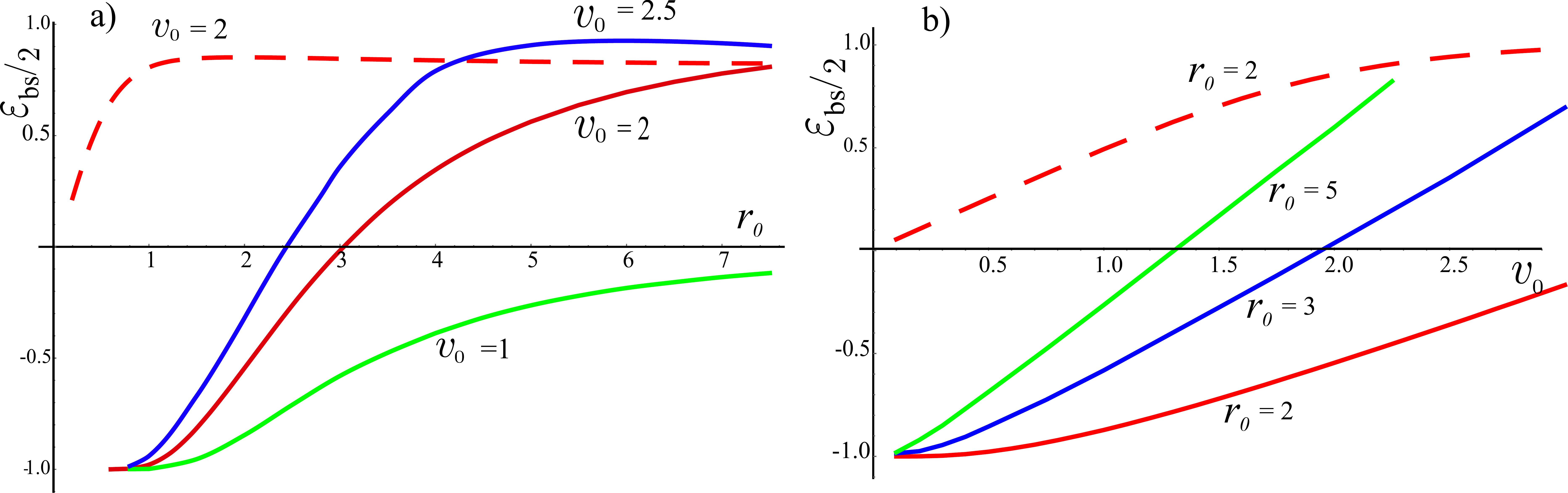}}
\caption{(Solid lines) Dependences of the ground state energy $\varepsilon_{bs}$ of the singlet BEPs of the first kind  in the topological phase: a) on the radius of the potential $r_0$ for various amplitudes $v_0$ (shown next to the lines); b) on the amplitude of  the potential  $v_0$ for various radii $r_0$ (shown next to the lines). Dashed lines show these dependences for BEPs of the second kind.}
\label{spectr}
\end{figure}

It was found that there are two kinds of the singlet bound states which differ in the composition of their band orbitals. In the BEPs of the first kind, both electrons are mainly in the same band, namely in the valence band for the repulsive interaction and in the conduction band for the attractive interaction. In the BEPs of the second kind both electrons are mainly in different bands. Figure~\ref{spectr} shows dependencies of the singlet BEP energy on the amplitude $v_0$ and radius $r_0$ of the interaction potential for both kind of states.

The wave function of the ground states of the singlet BEPs with the angular quantum number $m=0$ can be written as (the indexes m=0 are omitted):
\begin{equation}
\begin{split}
 \Psi_{bs}^{s}(r)&= C\left[\psi_{3}(r)\left(|E\uparrow E\downarrow\rangle-|E\downarrow E\uparrow\rangle\right)\right.\\
 &\left.+ \psi_{4}(r)e^{i\varphi}\left(|E\uparrow H\downarrow\rangle+|H\downarrow E\uparrow\rangle\right)\right.\\
 &\left.+ \psi_{7}(r)e^{-i\varphi}\left(|H\uparrow E\downarrow\rangle+|E\downarrow H\uparrow\rangle\right)\right.\\
 &\left.+ \psi_{8}(r)\left(|H\uparrow H\downarrow\rangle-|H\downarrow H\uparrow\rangle\right)\right],
\label{singlet_WF}
\end{split}
\end{equation}
where $C$ is a normalization constant.

\begin{figure}[htb]
\centerline{\includegraphics*[width=1\linewidth]{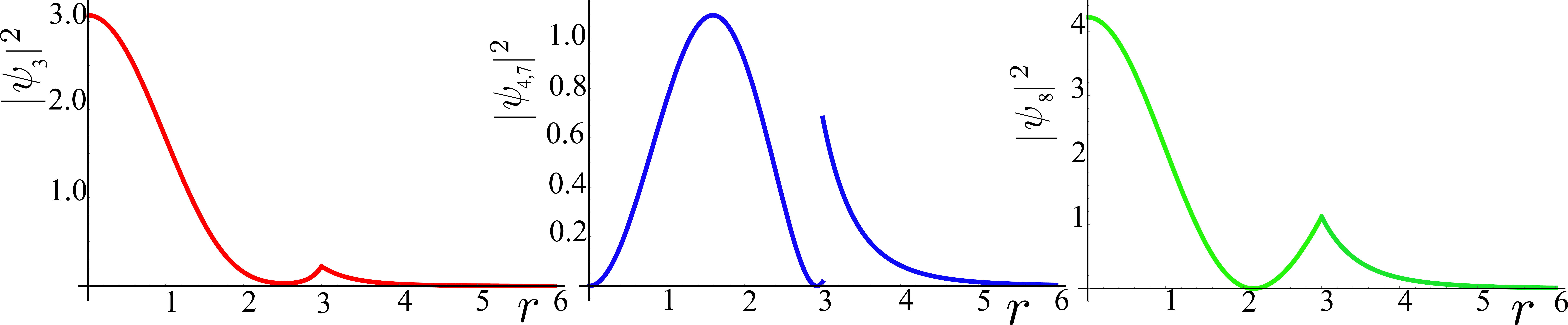}}
\caption{Radial dependence of the spinor components of the BEP of the first kind with parameters: energy $\varepsilon/2= 0.69$, $a=2.1$, $v_0=3.0$, $r_0=3.0$, $m=0$    in the topological phase .}
\label{sp_1}
\end{figure}

\begin{figure}[htb]
\centerline{\includegraphics*[width=1\linewidth]{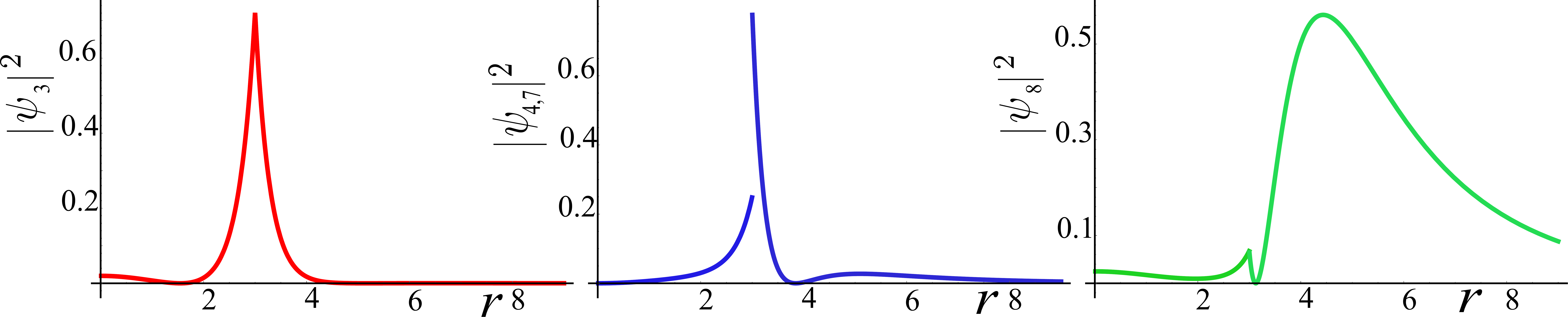}}
\caption{Radial dependence of the spinor components of the BEP of the second kind with parameters: energy $\varepsilon/2= 0.963$, $a=2.1$, $v_0=3.0$, $r_0=3.0$, $m=0$  in the topological phase.}
\label{sp_2}
\end{figure}

Figures~\ref{sp_1} and \ref{sp_2} show the results of numerical calculations of the envelope functions $\psi_{i}(r)$ of the singlet BEP spinors of the first and second kinds, respectively, with $m=0$. The wave functions of the singlet BEPs of both kinds consist of the same set of basis functions. But their envelope functions are different and the spatial distributions of the densities of two kinds of BEPs are different. The density of the BEP of the first kind is located mainly in the interaction region. The  density of the BEP of the second kind is distributed around the periphery of the interaction region. 

Now we turn to pairs of electrons that interact with each other, but do not form a bound state. There are two types of such states into which the BEP can decay: one with both electrons in the valence band and other with one electron in the valence band and other in the conduction band. First of all, we find the energies of these states. As the motion of the electrons forming a pair is not limited to any finite region of space, the energy can be defined by the asymptotics of the wave function at infinity where the interaction vanishes and the electrons can be considered as noninteracting. If there is no interaction, the momentum of each electron has a definite value $\mathbf{k}_{1,2}$ and the energy of the pair can be found as the sum of the single-particle energies defined by the Hamiltonian~(\ref{H_0})

\begin{equation}
\varepsilon_{\mathbf{k},\lambda}=\lambda \sqrt{(\tau+k^2)^2+a^2k^2}\,,
\end{equation}
where $\lambda$ is the band index, $\lambda=+$ relates to the conduction band and $\lambda=-$ to the valence band. $\tau$ is introduced to separate the topological ($\tau=-1$) and trivial ($\tau=1$) phases. The momenta $\mathbf{k}_{1,2}$ can be written in terms of the center-of-mass momentum $\mathbf{K}$ and the relative momentum $\mathbf{k}$,  $\mathbf{k}_{1,2}=\mathbf{k}\pm \mathbf{K}/2$. 

Thus, the energy of an unbound electron pair is determined by the momentum  $\mathbf{K}$, the relative momentum of free electrons $\mathbf{k}$ at infinity and indexes of the bands $\lambda_1,\lambda_2$:
\begin{multline}
\varepsilon_{\mathbf{K},\mathbf{k}}^{\lambda_1,\lambda_2}=\lambda_1\sqrt{\left(\tau+\left(\frac{\mathbf{K}}{2}+\mathbf{k}\right)\right)^2+a^2\left(\frac{\mathbf{K}}{2}+\mathbf{k}\right)^2}\\+\lambda_2\sqrt{\left(\tau+\left(\frac{\mathbf{K}}{2}-\mathbf{k}\right)\right)^2+a^2\left(\frac{\mathbf{K}}{2}-\mathbf{k}\right)^2}\,
\label{unbound_dispertion}
\end{multline}
and does not depend on the spin. But the wave function, of course, depends and will be also labeled  by the spin indexes $s_{1,2}$ of the electrons and be written as $\Phi_{\mathbf{K},\mathbf{k},s_1,s_2}^{\lambda_1,\lambda_2}$. 

\begin{figure}[htb]
\centerline{\includegraphics*[width=1\linewidth]{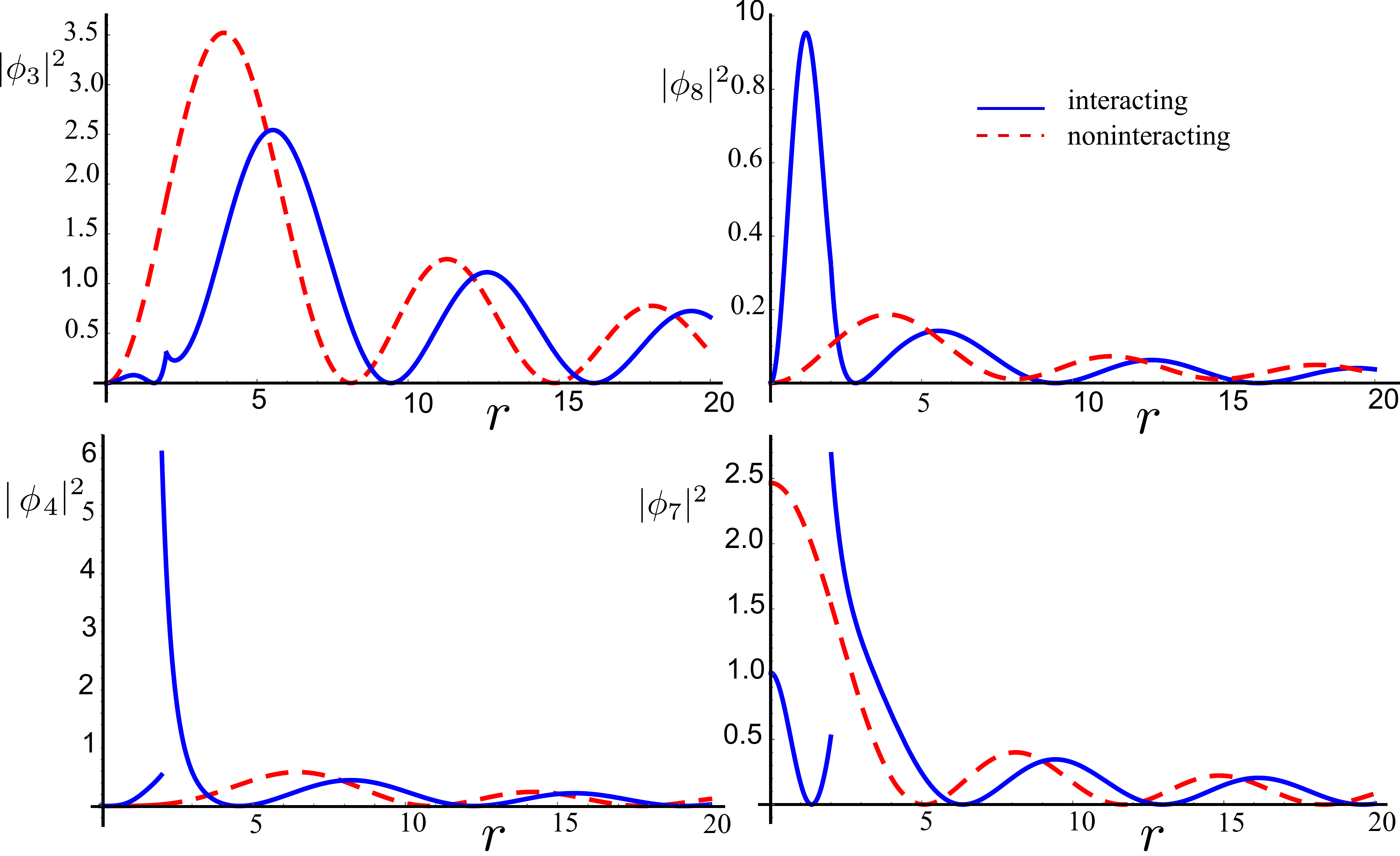}}
\caption{Spinor components of the unbound electron pair with $m=1$ as functions of the distance between electrons $r$ for interacting, $v_0=2$ (solid lines), and noninteracting, $v_0=0$ (dashed lines), electrons in topological phase. The parameters used in calculations are: energy $\varepsilon/2= -1.1$, $a=2.1$,  $r_0=2.0$.}
\label{fw_1}
\end{figure}

Details of the $\Phi_{\mathbf{K},\mathbf{k},\uparrow,\downarrow}^{v,v}$ calculation  are given in Appendix~\ref{App_wave_func}. Of interest are the components  with angular quantum number $m=1$, since the BEPs with angular quantum number $m=0$ decay into these states (see inthe following). The results of numerical calculation for the interacting and noninteracting pairs of unbound electrons with the same energy are shown in Fig.~\ref{fw_1}, where the spinor components $\phi_{i}(r)$ of the states with $m = 1$ are presented as functions of the relative position of the electrons. The total singlet two-particle wave function of unbound states $\Phi_{\mathbf{K},\mathbf{k}}^{v,v}$ that are used in the following for the calculation of the decay rates are found  antisymmetrizing the functions $\Phi_{\mathbf{K},\mathbf{k},\uparrow,\downarrow}^{v,v}$ with respect to the particle permutation.

\section{Radiative decay of the BEPs.}

To determine the radiative decay rate of BEPs we obtain the two-particle Hamiltonian $H_{int}(1,2)$ of the interaction of the pair of electrons with electromagnetic field within the electric dipole approximation by substitution $\mathbf{k}_{1,2}\to \mathbf{k}_{1,2}+(e/\hbar c)\mathbf{A}$ into the two-electron Hamiltonian of the BHZ model (see Appendix, B).

The decay rate is studied in the standard way using the Fermi’s Golden rule and considering $H_{int}(1,2)$ as a perturbation. An initial state is a BEP in a state $|bs\rangle=\Psi_{bs}(\mathbf{r}_1-\mathbf{r}_2)$ with an energy $\varepsilon_{bs}$ and an electromagnetic field in vacuum state $|\Omega\rangle$, and a final state is two electrons in one of the possible band states $|fs\rangle=\Phi_{\mathbf{K},\mathbf{k},s_1,s_2}^{\lambda_1,\lambda_2}$ with an energy $\varepsilon_{\mathbf{K},\mathbf{k}}^{\lambda_1,\lambda_2}$ and one photon with a wave vector $\mathbf{q}$, a polarization $\mathbf{e}_{\nu}$ and an energy $\varepsilon_q$. The total transition rate is obtained by summing over all possible final states: 

\begin{widetext}
\begin{equation}
\Gamma =\frac{2\pi|M|}{\hbar} \sum_{\shortstack{$\scriptstyle \lambda_1,s_1$\\ $\scriptstyle \lambda_2,s_2,\nu$}}\!\iiint \frac{d^3q}{(2\pi)^3} \frac{d^2k}{(2\pi)^2} \frac{d^2K}{(2\pi)^2} \Bigl|\langle fs|\otimes\langle \mathbf{q},\nu|H_{int}(1,2)|bs\rangle\otimes|\Omega\rangle\Bigr|^2 \delta(\varepsilon_{bs}-\varepsilon_{\mathbf{K},\mathbf{k}}^{\lambda_1,\lambda_2}-\varepsilon_q)\delta_{q_{||},K}\,,
\label{rate}
\end{equation}
\end{widetext}
where $q_{||}$ is an in-plain component of a light wave vector.

\begin{figure}[htb]
\centerline{\includegraphics*[width=0.8\linewidth]{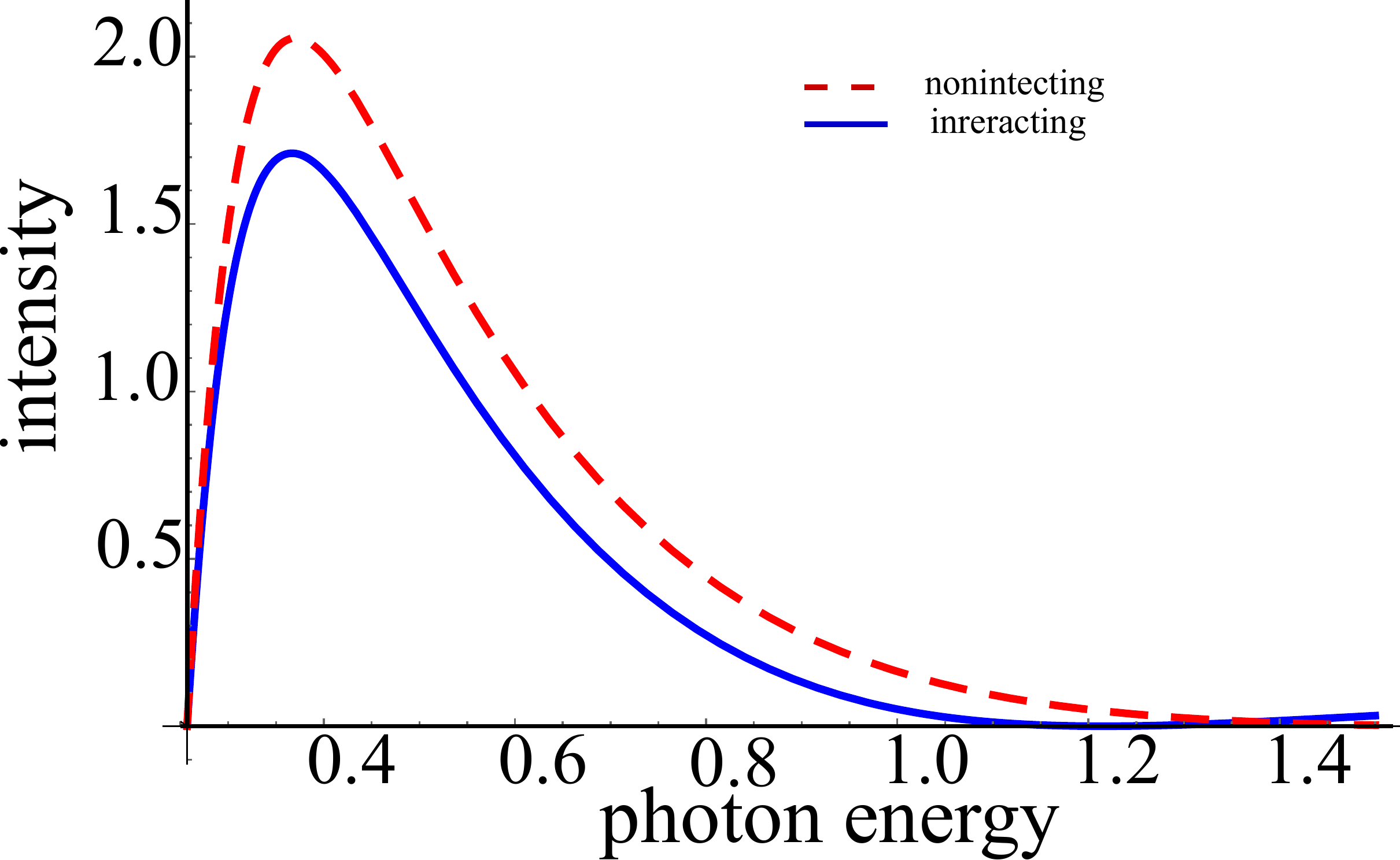}}
\caption{Radiative transition spectrum for the decay of the BEP of the first type into two unbound electrons in the valence band in topological phase. The BEP energy $\varepsilon_{bs}/2= -0.875, a = 2.1, r_0 = 2.0$. The wave functions of two-free-electron states are calculated with $v_0 = 0$ for the upper (dashed) and $v_0 = 1$ for the lower lines}. 
\label{radiation_spectr}
\end{figure}

We consider here a BEP in an empty crystal, that is, suppose that the valence band is not filled with electrons, and all states in the valence band are accessible to electrons when the pair decays.  This allows us to find an upper estimate for the decay rate, since the filling the bands leads to a decrease in the decay rate by about $(1-f)^2$ times, where $f$ is the filling factor. In addition, filling the bands leads to screening of the electron-electron interaction, electron correlation, and other effects that are beyond the scope of this article.

Two types of final states: one with two unbound electrons in the valence band, and the other with one electron in the conduction band and the other in the valence band, are taken into account by summing over $ \lambda_1, \lambda_2 $. The radiative decay of a singlet BEP with energy above the middle of the band gap is possible in both types of final states of unbound electrons, and we summarize these rates. However, these BEPs are not very interesting, since in the more general and realistic case of BHZ model with asymmetric electron and hole bands they are metastable~\cite{Sablikov1}. More stable BEPs with energies below the middle of the band gap can decay only into two electrons in the valence band due to energy conservation.

As a result of the conservation of the angular moment, the BEP in the ground state with $m = 0$ decays into components  of a final two electron spinor with an angular moment $m=\pm1$ (see Appendix, B). Since the final two-electron states have a continuous spectrum, a radiative decay of a BEP with an energy $\varepsilon_{bs}$ is allowed into two electrons in the valence band with the emission of photons with an energy in a wide range $\varepsilon_q > \varepsilon_{bs} +2$. Therefore, the spectrum of radiative transitions is very broad, as shown in Fig.~\ref{radiation_spectr}. The shape of these curves is mainly determined by two factors. The first is the overlap integral of the  BEP wave function and the wave function of the final pair of free electrons, which decreases rapidly with increasing momentum of free electrons. And the second is the density of the final electron states and the phase space available for the emission of photons, which increase with increasing energy.

\begin{figure}[htb]
\centerline{\includegraphics*[width=1\linewidth]{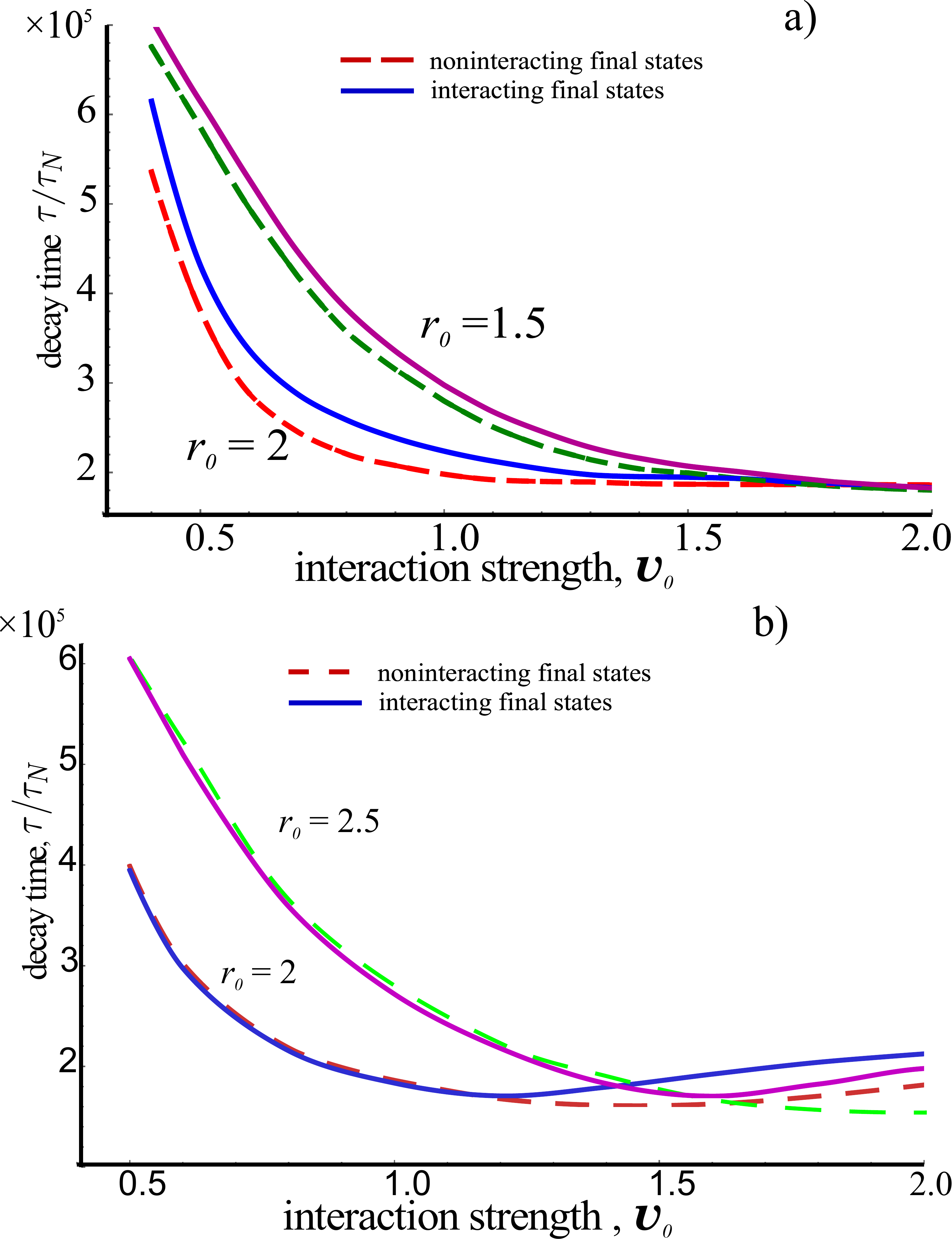}}
\caption{Decay time of the BEP of the first type versus the interaction potential amplitude in topological phase a) and trivial phase b). The interaction radii are shown next to the lines. Solid lines are calculated taking into account the e-e interaction in final states, when the dashed lines are calculated neglecting the e-e interaction in final states.}
\label{decay_int}
\end{figure}

 We have studied how the e-e interaction in the final state affects the decay time of BEPs in the topological and trivial phases, and found that the effect is different for these phases. The results are shown in Fig.~\ref{decay_int}. In the topological phase, the e-e interaction leads to an increase in the decay time when the interaction is weak (in Fig.~\ref{decay_int}a for $v_0 < 1.5$ ). However, the effect becomes opposite, albeit substantially weaker, at strong e-e interaction. In the trivial phase, on the contrary, the decay time decreases very weakly with the e-e interaction at small $v_0$, $v_0 < 1.5 $ in Fig.~\ref{decay_int}b, and  increases significantly when the interaction is strong.

The effect of the e-e interaction in the final state on the decay time depends on whether the interaction leads to an increase in the wave function of the unbound pair in the region of small distances, where the interaction acts and the bound state is localized, or to its decrease. In other words, the effect depends on whether the unbound electrons attract or repel each other. Attraction leads to a decrease in decay time, and repulsion leads to an increase. It is clear that attraction occurs at a negative reduced effective mass of electrons, and repulsion at a positive one. In turn, the sign of the reduced effective mass is determined by mixing of the states of the electron and hole bands in the pair state under given conditions. Thus, the different effect of e-e interaction in the topological and trivial phases, as well as in the case of strong and weak interactions, is due to the different contributions of the electron and hole bands to the spinor of an electron pair.

\begin{figure}
\centerline{\includegraphics*[width=1\linewidth]{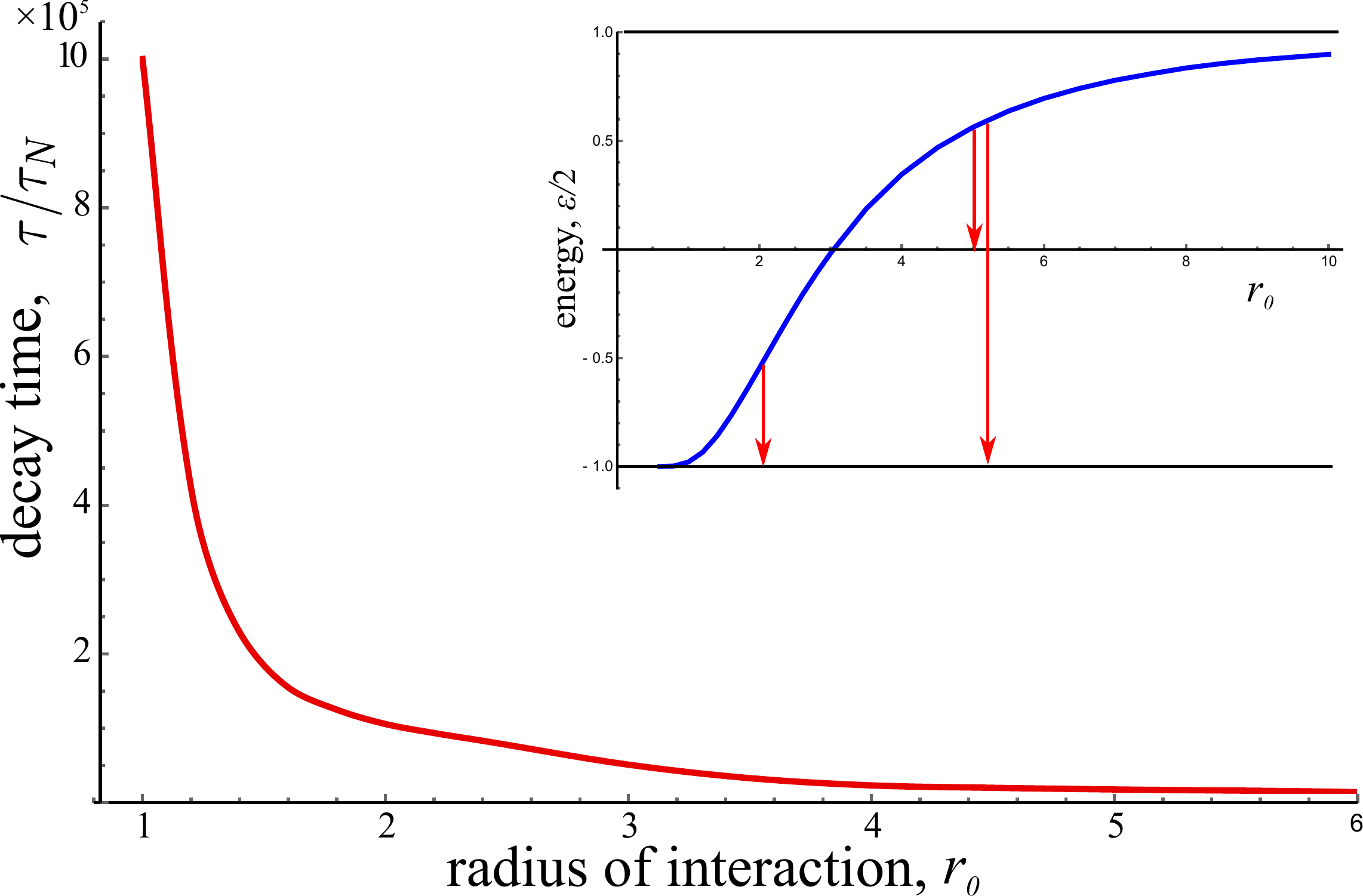}}
\caption{Decay time of the BEP of the first type in the TI versus the interaction potential radius. Insert shows the spectrum and allowed decay channels. The parameters used in the calculations are $a = 2.1$, $v_0 = 2.0$, $m = 0$.}
\label{decay_first}
\end{figure}
 
Figure~\ref{decay_first} shows the radiative decay time versus the interaction potential radius for the BEPs of the first kind in the TI. The lifetime of BEP decreases in order of magnitude with increasing of the BEP energy mainly due to the increase of density of the photon states with increasing photon energy as well as appearing the second decay channel for BEPs with energies higher the middle of bandgap.

Numerical estimate of the decay time for parameters close to those of quantum well of thickness $d=7$~nm on the heterostructures HgTe/CdHgTe with BHZ parameters  $ A$ = 3.65~eV~\AA, $M = -0.010$~eV, $ B$ = - 68.6~eV~\AA$^2$ is as follows. The normalizing time $\tau_N=\kappa \hbar/(4\pi^2e^2)(|B/M|)^{1/2}$ is estimated as $\tau_N \approx 2\cdot 10^{-14}$~s and the decay time is $\tau \sim 10^{-9}$~s. It is obvious, that the decay time of BEPs is  extremely large than the radiative decay time of excitons in direct-gap semiconductors. But it is comparable with lifetime of indirect excitons, which is long due to the spatial separation of electrons and holes~\cite{Suris}.

Such a long  BEP decay time is mainly due to the structure of the two-electron wave functions in the initial and final states. First, the wave function of the initial and final states weakly overlap, because in the initial state the electrons are localized near each other and in the final state they are free propagating. Second, the spinor components of both states have different signs, so some terms in  transition matrix elements partially cancel each other. Therefore although the BEPs can decay  by radiation of photons with energies in a broad range,  the radiative life time of BEP is large due to a very low oscillator strength.  

In Fig.~4 of Ref.~\onlinecite{Sablikov1} the decay times of BEPs in topological and trivial phases were presented for various model parameters. The curves were calculated for the BEP states with close dependencies of their energies on the interaction potential $v_0$. Thus the densities of phase space for the photon emission  where also close. Calculations accounting the e-e interaction in the final  states do not change those results qualitatively.  Namely, the BEPs in the topological phase with  $a=\sqrt{2}$ and nearly flat bands in the extrema, have the largest decay time. In the topological phase with $a > 2$ and a quadratic spectrum in the extrema, the decay time is much shorter than in the former case. The large difference of decay times for these states is mainly a result of different densities of final electron states. In the trivial phase the BEP life time is much shorter than in the topological phase, with other things being equal. These results show that  both the topological properties of the electronic states and the band dispersion near the extrema play an important role in a stability of the BEPs,  though the e-e interaction affects also the BEPs decay time.

Triplet BEPs with energies above the middle of the bandgap decay very slowly into pairs in which one of the electrons is in the conduction band and the other in the valence band. But their decay into pairs of electrons in the valence band are not allowed. For the singlet BEPs the latter is the main decay channel.  For the triplet BEPs the transition amplitudes of the spinor components for transition into a pair of electrons in a valence band exactly cancel  each other. Therefore, triplet BEPs with energies below the middle of the bandgap are dark. 

\section{Conclusion}

We studied the radiative decay of the BEPs in two-dimensional materials described by the BHZ model. BEP lifetime is one of the main parameters that can determine the role of BEPs in nonequilibrium and collective processes in topologically non-trivial materials. Strongly bound triplet BEPs with energies below the middle of the band gap are dark: in the second order of perturbation theory, they cannot decay with the emission of photons. The radiative decay time  of bright singlet BEPs is quite large in the scale of characteristic relaxation times of the electron system. Under real conditions of the HgTe/CdHgTe heterostructures, the decay time is estimated at the nanosecond level. 

A rather long decay time of BEPs compared to the decay time of 2D direct excitons  results from two factors: small oscillator strength  due to weak overlap of the wave function of the final free electrons and the wave function of the BEPs, and the restriction  of the phase space, where the radiative transition is possible, imposed by the requirement of the energy and momentum conservation. 

We explored the effect of inter particle interaction on the wave functions of the pair of unbound electrons  to find out how it changes the  BEPs decay time. The e-e interaction noticeably affects the wave functions of unbound electron pairs and, therefore the radiative decay time. The account of e-e interaction of free electrons in topological phase results in a noticeably increase of the BEPs decay time when the interaction is not very strong. But when the interaction is large enough to change the composition of the spinor of a final state, the interaction can reduce the decay time. The  values of the BEP decay times are of the same order as to those in Ref.~\onlinecite{Sablikov1}. 

The topology of the band states as well as the band dispersion strongly affect the radiative decay rate: the decay time in the topological phase is essentially longer than in the trivial phase. Moreover, the dependence of the BEP decay time on interaction strength is different in two phases. In these phases the two-particle wave functions  have very different spinor structure. In the trivial phase, the dominant component of the spinor is that corresponding the configuration with both electrons in the hole band. Hence the matrix element of the transition of electrons to the valence band is relatively large. On the contrary, in the topological phase, all components of the spinor are of the same order. 

The terms in the BEP--light interaction Hamiltonian,  which determine the oscillator strength of BEPs, are proportional the band hybridization parameter $a$. This parameter determines not only  the matrix element of the optical transition, but also the band dispersion; therefore, the decay time substantially depends on $ a $. For $a=\sqrt{2}$ the dispersion of the bands in the topological phase is almost flat. In this case the BEPs has the largest binding energy and the longest decay time. It can be expected that BEPs in such materials are the most stable. 

It is important to note that we estimated the lower limit of BEP lifetime assuming that the final electron states in the bands are not occupied. Due to Pauli blocking of transitions to occupied states, it can be expected that the decay time will be even longer if we take into account the filling of the band states. The filling of the valence band results not only in reducing the states available for BEPs decay, but also in  screening of the e-e interaction, electron correlation  and other effects that are beyond the scope of this article.

To know the decay time of the BEPs in real structures a study of the nonradiative decay and many-particle effects are required. Nevertheless, our study  of the radiative decay time of BEPs shows a  possibility of  their  manifestations in transport and collective processes in topologically nontrivial materials.

\acknowledgments
This work was carried out in the framework of the state task and partially supported by the Russian Foundation for Basic Research, project No. 20-02-00126

\appendix
\section{Wave functions}\label{App_wave_func}

The spinor components of the singlet ${\Psi}_s(\mathbf{r})$ (the index $\mathbf{K}=0$ is dropped hereinafter) are defined by:
\begin{equation}\label{singlet_P2}
\left\{
\begin{array}{rl}
 \left[2v(r)\!-\!\varepsilon\!+\!2\tau\!+\!2\hat{k}^2\right]\psi_3(\mathbf{r})\!+\!a\hat{k}_-\psi_4(\mathbf{r})\!+\!a\hat{k}_+\psi_7(\mathbf{r}) &\!=0\\
 a\hat{k}_+\psi_3(\mathbf{r})+\left[2v(r)-\varepsilon\right]\psi_4(\mathbf{r})+a\hat{k}_+\psi_8(\mathbf{r}) &\!=0\\
 a\hat{k}_-\psi_3(\mathbf{r})+\left[2v(r)-\varepsilon\right]\psi_7(\mathbf{r})+a\hat{k}_-\psi_8(\mathbf{r}) &\!=0\\
 a\hat{k}_-\psi_4(\mathbf{r})\!+\!a\hat{k}_+\psi_7(\mathbf{r})\!+\!\left[2v(r)\!-\!\varepsilon\!-\!2\tau\!-\!2\hat{k}^2\right]\psi_8(\mathbf{r}) &\!=0\,.
\end{array}\right. 
\end{equation}

It is convenient to go to polar coordinates ($r,\varphi$) and write the wave functions as:
\begin{equation}\label{Psi2_F}
 \Psi_s(\mathbf{r})\!=\!\sum\limits_m \Psi_{sm} e^{im\varphi}\!=\!\sum\limits_m
 \begin{pmatrix}
 \psi_{3m}(r) \\ \psi_{4m}(r) e^{i\varphi} \\ \psi_{7m}(r) e^{-i\varphi} \\ \psi_{8m}(r)
 \end{pmatrix}
 e^{im\varphi}.
\end{equation} 

The system of Eqs~(\ref{singlet_P2}) reduces to independent systems of four equations defining the components $\psi_{3m}, \psi_{4m}, \psi_{7m}$, and $\psi_{8m}$ for each $m$. 

Matching conditions for the wave functions at the radius $r=r_0$ of the steplike potential can be obtained by integrating the  equations over the transition region, $|r-r_0|<\delta$, assuming that $v(r)$ is a finite value and taking the limit $\delta\to 0$. Thus we obtained the following matching equations for singlet-like states:

\begin{equation}
 \begin{array}{rl}\label{match_cond}
 \psi_{3m}\bigm|_-^+&=0,\\
 \psi_{8m}\bigm|_-^+&=0,\\
 \frac{d\psi_{3m}}{dr}+ia(\psi_{4m}+\psi_{7m})\biggm|_-^+&=0,\\
 \frac{d\psi_{3m}}{dr}+\frac{d\psi_{8m}}{dr}\biggm|_-^+&=0.
\end{array}
\end{equation} 

The wave functions $\Psi_s(\mathbf{r})$  are determined solving Eqs.~(\ref{singlet_P2}) in two regions $r<r_0$ and $r>r_0$ and then matching the found functions at $r=r_0$ using Eqs.~(\ref{match_cond}).

Here we present the calculation details for topological phase with $\tau = -1 $.  

In the case of the step potential $v(r)$ , Eqs.~(\ref{singlet_P2}) are easily solved in terms of the Bessel functions. The fundamental set of solutions for the components of the spinor $\Psi_s$ has the form:
\begin{equation}\label{Bessel_base}
 \begin{array}{ll}
  \psi_{3m}(r)=A^{\pm}_m \mathcal{F}_m(Q_{\pm}r), &\psi_{4m}(r)=B^{\pm}_m \mathcal{F}_{m+1}(Q_{\pm}r),\\ \psi_{7m}(r)=C^{\pm}_m \mathcal{F}_{m-1}(Q_{\pm}jr), &\psi_{8m}(r)=D^{\pm}_m \mathcal{F}_m(Q_{\pm}r),
 \end{array}
\end{equation} 
where the wave numbers $Q_{\pm}$ are the roots of the dispersion equation, which has a unified form in both regions:
\begin{equation}
 \widetilde{\varepsilon}[\widetilde{\varepsilon}^2-(1-Q^2)^2-a^2Q^2]=0,
\end{equation} 
where $\widetilde{\varepsilon}$ takes different values for the interaction region and the outer region,
\begin{equation}
 \widetilde{\varepsilon}=\left\{
 \begin{array}{ll}
  \varepsilon_0 - v_0\,,\; & r<r_0,\\
  \varepsilon_0\,,\; & r>r_0.
 \end{array}
 \right.
\end{equation} 
We have denoted here $\varepsilon_0\equiv \varepsilon/2$, which is the energy of an electron pair per particle. 

In Eq.~(\ref{Bessel_base}) $\mathcal{F}_m(Q_{\pm}r)$ can be written as any pair of the Bessel functions~\cite{Sablikov}, the choice of which in a specific case is determined by the values of $Q_{\pm}$ at given $a$ and $\widetilde{\varepsilon}$, and by the behavior of the Bessel function at $r\to 0$ and $r\to \infty$. We consider here $a^2 > 4$. 

In the energy interval $-1<\varepsilon_0<-1+v_0$, the solution of Eqs~(\ref{singlet_P2}) for bound states can be presented in the following form:\\

(i) at $r<r_0$,
\begin{equation}\label{wave_fun1}
 \begin{array}{ll}
 \psi_{3m}=& A_+J_m(k_+r)+A_-I_m(k_-r),\\
 \psi_{4m}=& A_+\mathcal{B}_+J_{m+1}(k_+r)+A_-\mathcal{B}_-I_{m+1}(k_-r),\\
 \psi_{7m}=& A_+\mathcal{C}_+J_{m-1}(k_+r)+A_-\mathcal{C}_-I_{m-1}(k_-r),\\
 \psi_{8m}=& A_+\mathcal{D}_+J_m(k_+r)+A_-\mathcal{D}_-I_m(k_-r),
 \end{array}
\end{equation} 
where
\begin{equation}
 \begin{array}{ll}
 \mathcal{B}_{\pm}=& i\dfrac{\varepsilon_0-v_0+1\mp k_\pm^2}{ak_\pm},\\
 \mathcal{C}_{\pm}=& -i\dfrac{\varepsilon_0-v_0+1\mp k_\pm^2}{ak_\pm},\\
 \mathcal{D}_{\pm}=& \dfrac{\varepsilon_0-v_0+1\mp k_\pm^2}{\varepsilon_0-v_0+1\pm k_\pm^2},
 \end{array}
\end{equation} 
and 
\begin{equation}\label{k_pm}
 k_{\pm}=\sqrt{\pm\!\left(1\!-\!\frac{a^2}{2}\right)\!+\!\sqrt{a^2\left(\frac{a^2}{4}\!-\!1\right)+(\varepsilon_0\!-\!v_0)^2}};
\end{equation} 

(ii) at $r>r_0$,
\begin{equation}\label{wave_fun2}
 \begin{array}{ll}
 \psi_{3m}=& B_+K_m(\kappa_+r)+B_-K_m(\kappa_-r),\\
 \psi_{4m}=& B_+\mathcal{K}_+K_{m+1}(\kappa_+r)+B_-\mathcal{K}_-K_{m+1}(\kappa_-r),\\
 \psi_{7m}=& B_+\mathcal{L}_+K_{m-1}(\kappa_+r)+B_-\mathcal{L}_-K_{m-1}(\kappa_-r),\\
 \psi_{8m}=& B_+\mathcal{M}_+K_m(\kappa_+r)+B_-\mathcal{M}_-K_m(\kappa_-r),
 \end{array}
\end{equation} 
where
\begin{equation}
 \begin{array}{rl}
 \mathcal{K}_{\pm}=\mathcal{L}_{\pm}&=-i\dfrac{\varepsilon_0+1+\kappa_\pm^2}{a\kappa_\pm},\\
 
 \mathcal{M}_{\pm}=& \dfrac{\varepsilon_0+1+\kappa_\pm^2}{\varepsilon_0-1-\kappa_\pm^2},
 \end{array}
\end{equation}
and
\begin{equation}
 \kappa_{\pm}=\sqrt{-1+\frac{a^2}{2}\pm \sqrt{a^2\left(\frac{a^2}{4}-1\right)+\varepsilon_0^2}}.
\end{equation} 

The functions~(\ref{wave_fun1}) and (\ref{wave_fun2}) should be matched at the boundary $r=r_0$. Using Eqs.~(\ref{match_cond}), we get a homogeneous system of equations  for the coefficients $A_+, A_-, B_+, B_-$.  The eigenenergies are determined by the equation 
\begin{equation}\label{DD}
\mathfrak{D}(\varepsilon_0;a,v_0,r_0,m)=0,
 \end{equation}
where $\mathfrak{D}$ is the determinant  of this equation system, which is a function of the energy $\varepsilon_0$ and the parameters $a, v_0, r_0, m$.
The calculated energy spectrum and eigenfunctions of bound states are shown in Figs.~\ref{spectr},\ref{sp_1},\ref{sp_2}. After antisymmetrizing of  the functions $\Phi_{s}$ with respect to particle permutation we get the wave function of the ground state of BEPs, Eq.(~\ref{singlet_WF}).

The wave functions of the unbound pairs in the presence of interaction can be calculated similarly the wave functions of the bound states. The only difference is that now the energy is a given quantity defined by Eq.~(\ref{unbound_dispertion}) and the wave function does not vanish at infinity. The spinor components of a singlet state with two electrons in the valence band are defined by Eqs.~(\ref{singlet_P2}), in which the energy is a value defined by Eq.~(\ref{unbound_dispertion}) and should be taken in the energy interval $\varepsilon_0<-1$. For clarity we denote the wave function of an unbound pair of electrons as $\Phi_{\mathbf{K=0},\mathbf{k},\uparrow,\downarrow}^{v,v}$ and its spinor components as $\phi_{im}$.
             
In the range of $r<r_0$ , the solution of Eqs.~(\ref{singlet_P2}) can be presented in the same form as Eqs.~(\ref{wave_fun1}) with replacement of $\psi_{im}$ by $\phi_{im}$ . In the range of $r>r_0$, the spinor components of the wave functions $\Phi_{\mathbf{K=0},\mathbf{k},\uparrow,\downarrow}^{v,v}$ can be written as:\\
\begin{equation}\label{func_free}
 \begin{array}{ll}
 \phi_{3m}=& G_+J_m(\kappa_+r)+C Y_m(\kappa_+r)+G_-K_m(\kappa_-r),\\
 \phi_{4m}=& G_+\mathcal{K}_+J_m(\kappa_+r)+C\mathcal{K}_+ Y_m(\kappa_+r)+G_-\mathcal{K}_-K_{m+1}(\kappa_-r),\\
 \phi_{7m}=& G_+\mathcal{L}_+J_m(\kappa_+r)+C\mathcal{L}_+Y_m(k_+r)+G_-\mathcal{L}_-K_{m-1}(\kappa_-r),\\
 \phi_{8m}=& G_+\mathcal{M}_+J_m(\kappa_+r)+C\mathcal{M}_+ Y_m(k_+r)+G_-\mathcal{M}_-K_m(\kappa_-r),
 \end{array}
\end{equation} 

Using Eqs.~(\ref{match_cond},\ref{wave_fun1}) with replacement of $\psi_{im}$ by $\phi_{im}$ and Eq.(\ref{func_free}), we get a system of equations for the coefficients $A_+, A_-, G_+, G_-, C$. We express these coefficients in terms of $G_+$, then define the coefficient $G_+$ from a normalization condition for the wave function. For a small $\mathbf{K}\ll 1$ we set $\Phi_{\mathbf{K},\mathbf{k},\uparrow,\downarrow}^{v,v}=\Phi_{\mathbf{K=0},\mathbf{k},\uparrow,\downarrow}^{v,v}e^{i\mathbf{KR}}$.

\section{Interaction Hamiltonian of a pair of electrons with an electromagnetic field}\label{App_EM-Ham}

To get a Hamiltonian of the interaction of a pair of 2D electrons with an in plane component of electromagnetic field $\mathbf{A}(r)=\mathbf{A}_\mathbf{g}\mathbf{e}_{\nu}e^{i\mathbf{g}\mathbf{r}}$ we do the standard replacement $\hat{k} \Rightarrow \hat{k}+(e/hc)\mathbf{A}$ in the one-particle Hamiltonian~(\ref{H_0}) and get $$\hat{H}(\mathbf{\hat{k}},\mathbf{A})= H_0(\mathbf{\hat{k}})+ \hat{H}_{int}(\mathbf{\hat{k}},\mathbf{A}),$$ where:
\begin{equation*}
 \hat{H}_{int}(\mathbf{\hat{k}},\mathbf{A})=
\begin{pmatrix}
 \hat{h_i}(\mathbf{\hat{k},A}) & 0\\
 0 & \hat{h_i}^*(\mathbf{-\hat{k}},-\mathbf{A})
\end{pmatrix}.
\end{equation*}
Here
\begin{equation*}
\hat{h_i}(\mathbf{\hat{k}},\mathbf{A})=
\begin{pmatrix}
 2\mathbf{\hat{k}\cdot A}\quad & a\mathbf{A}_{+}\\
 a\mathbf{A_{-}} & -2\mathbf{\hat{k}\cdot A}
\end{pmatrix}.
\end{equation*}
The interaction of a two-particle state with an electromagnetic field is described by the two-particle Hamiltonian
\begin{equation*}
 \hat{H}_{int}(1,2)=\hat{H}_{int}(\mathbf{\hat{k}}_1,\mathbf{A}(\mathbf{r}_1))\oplus\hat{H}_{int}(\mathbf{\hat{k}}_2,\mathbf{A}(\mathbf{r}_2))\,.
\end{equation*}
For a field with a wavelength much larger than a BEP radius, $|\mathbf{gr}_b|\ll 1$, where $\mathbf{g}$ is the light wave vector and  $\mathbf{r}_b$ is the characteristic BEP radius, the BEP-field interaction Hamiltonian can be written as:
\begin{equation}
 \hat{H}_{int}(\mathbf{\hat{k},A})=e^{i\mathbf{gR}}
\begin{pmatrix}
 H_{11}\quad & H_{12}\quad&0&0\\
 H_{21} & H_{22}&0&0\\
 0&0&H_{11} & -H_{21}\\
 0&0&-H_{12} & H_{22}\\
\end{pmatrix}.
\label{H_int}
\end{equation}
Here $H_{12}=H_{21}^* =a\mathbf{A_+}\cdot\mathbf{I}_{4\times 4}\,\,$, and
\begin{equation*}
   H_{11}=
\begin{pmatrix}
  2\mathbf{\hat{K}\cdot A} & a\mathbf{A_+}&0&0\\
 a\mathbf{A_-} & -4\mathbf{\hat{k}\cdot A}\quad&0&0\\
 0&0&-2\mathbf{\hat{K}\cdot A} & -a\mathbf{A_-}\\
 0&0&-a\mathbf{A_+} & 4\mathbf{\hat{k}\cdot A}\quad\\
\end{pmatrix},
\end{equation*}

\begin{equation*}
  H_{22}=
\begin{pmatrix}
 4\mathbf{\hat{k}\cdot A}\quad & a\mathbf{A_{+}}&0&0\\
 a\mathbf{A_{-}} &-2\mathbf{\hat{K}\cdot A}&0&0\\
 0&0&-4\mathbf{\hat{k}\cdot A}\quad & -a\mathbf{A_{-}}\\
 0&0&-a\mathbf{A_{+}} & 2\mathbf{\hat{K}\cdot A}\\
\end{pmatrix}.
\end{equation*}

In addition to usual term $\mathbf{k}\mathbf{A}$ the Hamiltonian~(\ref{H_int})  contains terms proportional to $A_{\pm}$, that result from the hybridization of the $e$- and $h$-bands and cause the decay of BEPs (see below). 

In the dimensionless form the vector potential $\mathbf{A}$ of vacuum fluctuations due to which the photon emission takes place is
\begin{equation}
\mathbf{A}(\mathbf{r,t})=\sum_{\mathbf{q},\nu} \sqrt{\frac{2\pi e^2|B|}{V\kappa M^2 \varepsilon_{\mathbf{q}}}} \mathbf{e}_{\nu} a_{\mathbf{q},\nu}^{\dag}e^{i(\varepsilon_\mathbf{q}t+\mathbf{q_{||}}\mathbf{r}+q_\bot z)},
\end{equation} 
where $\kappa$ is the dielectric constant of the material, $V$ is a normalization volume, $\varepsilon_q=\hbar\omega_q/|M|$ is a photon energy, $\mathbf{q}$ is a photon wave vector, and $\mathbf{e}_{\nu}$ is a polarization vector.
For the light with the wave vector $\mathbf{q}=(\mathbf{q_{||}},q_\bot)$ propagating along an arbitrary direction, $\mathbf{q} = \mathbf{n}\sqrt{q_{||}^2+q_\bot^2}$, where $\mathbf{n} =(\sin{\theta}\cos{\beta}, \sin{\theta}\sin{\beta}, \cos{\theta})$, the in-plane components of the vector-potential $\mathbf{A}$ contributing to the emission process are:

\begin{equation*}
\begin{split}
 \mathbf{A_{+}}=&A_x+\!i A_y=\frac{1}{2}A_q e^{i\beta}(\cos\theta \pm 1)e^{i\varepsilon_\mathbf{q} t}\\
 \mathbf{A_{-}}=&A_x-\!i A_y=\frac{1}{2}A_q e^{-i\beta}(\cos\theta \mp 1)e^{i\varepsilon_\mathbf{q} t}.
\end{split}
\end{equation*}
Here the upper and lower signs are for right- and left-circular polarization respectively.
The transition amplitudes are defined as 
\begin{widetext}
\begin{equation}
\begin{split}
\langle fs|\otimes\langle \mathbf{q},\nu|H_{int}(1,2)|bs\rangle\otimes|\Omega\rangle\ =
\int d^2R  e^{-i(\mathbf{K}-\mathbf{q}_{||})\mathbf{R}}\int_0^\infty r dr\int_0^{2\pi} d\varphi\sum_m[1-(-1)^m]\times \\
\left\{(\phi_{3m}(r)-\phi_{8m}(r)) (a A_+\psi_{70}(r)e^{-i(m-1)\varphi}-a A_-\psi_{40}(r)e^{-i(m+1)\varphi})+\right.\\
\left. (\psi_{30}(r)-\psi_{80}(r)) (a A_-\phi_{7m}(r)e^{-i(m-1)\varphi}- a A_+\phi_{4m}(r)e^{-i(m+1)\varphi})\right\} \\ 
= 8\pi^2\delta(\mathbf{K}-\mathbf{q}_{||})\int_0^\infty r dr\sum_m\left\{(\phi_{3m}(r)-\phi_{8m}(r))(a A_+\psi_{70}(r)\delta_{m,-1}- \right. \\
\left. a A_-\psi_{40}\delta_{m,1})+(\psi_{30}(r)-\psi_{80}(r))(a A_-\phi_{7m}(r)\delta_{m,1}- a A_+\phi_{4m}(r)\delta_{m,-1})\right\}
\end{split}
\end{equation}
\end{widetext}

\bibliography{BEP_radiative_decay}

\begin{thebibliography}{22}%
\makeatletter
\providecommand \@ifxundefined [1]{%
 \@ifx{#1\undefined}
}%
\providecommand \@ifnum [1]{%
 \ifnum #1\expandafter \@firstoftwo
 \else \expandafter \@secondoftwo
 \fi
}%
\providecommand \@ifx [1]{%
 \ifx #1\expandafter \@firstoftwo
 \else \expandafter \@secondoftwo
 \fi
}%
\providecommand \natexlab [1]{#1}%
\providecommand \enquote  [1]{``#1''}%
\providecommand \bibnamefont  [1]{#1}%
\providecommand \bibfnamefont [1]{#1}%
\providecommand \citenamefont [1]{#1}%
\providecommand \href@noop [0]{\@secondoftwo}%
\providecommand \href [0]{\begingroup \@sanitize@url \@href}%
\providecommand \@href[1]{\@@startlink{#1}\@@href}%
\providecommand \@@href[1]{\endgroup#1\@@endlink}%
\providecommand \@sanitize@url [0]{\catcode `\\12\catcode `\$12\catcode
  `\&12\catcode `\#12\catcode `\^12\catcode `\_12\catcode `\%12\relax}%
\providecommand \@@startlink[1]{}%
\providecommand \@@endlink[0]{}%
\providecommand \url  [0]{\begingroup\@sanitize@url \@url }%
\providecommand \@url [1]{\endgroup\@href {#1}{\urlprefix }}%
\providecommand \urlprefix  [0]{URL }%
\providecommand \Eprint [0]{\href }%
\providecommand \doibase [0]{http://dx.doi.org/}%
\providecommand \selectlanguage [0]{\@gobble}%
\providecommand \bibinfo  [0]{\@secondoftwo}%
\providecommand \bibfield  [0]{\@secondoftwo}%
\providecommand \translation [1]{[#1]}%
\providecommand \BibitemOpen [0]{}%
\providecommand \bibitemStop [0]{}%
\providecommand \bibitemNoStop [0]{.\EOS\space}%
\providecommand \EOS [0]{\spacefactor3000\relax}%
\providecommand \BibitemShut  [1]{\csname bibitem#1\endcsname}%
\let\auto@bib@innerbib\@empty
\bibitem [{\citenamefont {Combescot}\ and\ \citenamefont
  {Shiau}(2015)}]{combescot2015excitons}%
  \BibitemOpen
  \bibfield  {author} {\bibinfo {author} {\bibfnamefont {M.}~\bibnamefont
  {Combescot}}\ and\ \bibinfo {author} {\bibfnamefont {S.-Y.}\ \bibnamefont
  {Shiau}},\ }\href {https://doi.org/10.1093/acprof:oso/9780198753735.001.0001}
  {\emph {\bibinfo {title} {Excitons and Cooper pairs: two composite bosons in
  many-body physics}}}\ (\bibinfo  {publisher} {Oxford University Press},\
  \bibinfo {year} {2015})\BibitemShut {NoStop}%
\bibitem [{\citenamefont {Kagan}(2013)}]{kagan2013modern}%
  \BibitemOpen
  \bibfield  {author} {\bibinfo {author} {\bibfnamefont {M.~Y.}\ \bibnamefont
  {Kagan}},\ }\href {https://doi.org/10.1007/978-94-007-6961-8} {\emph
  {\bibinfo {title} {Modern trends in superconductivity and superfluidity}}},\
  \bibinfo {series} {Lecture Notes in Physics}, Vol.\ \bibinfo {volume} {874}\
  (\bibinfo  {publisher} {Springer},\ \bibinfo {year} {2013})\BibitemShut
  {NoStop}%
\bibitem [{\citenamefont {Micnas}\ \emph {et~al.}(1990)\citenamefont {Micnas},
  \citenamefont {Ranninger},\ and\ \citenamefont
  {Robaszkiewicz}}]{RevModPhys.62.113}%
  \BibitemOpen
  \bibfield  {author} {\bibinfo {author} {\bibfnamefont {R.}~\bibnamefont
  {Micnas}}, \bibinfo {author} {\bibfnamefont {J.}~\bibnamefont {Ranninger}}, \
  and\ \bibinfo {author} {\bibfnamefont {S.}~\bibnamefont {Robaszkiewicz}},\
  }\href {\doibase 10.1103/RevModPhys.62.113} {\bibfield  {journal} {\bibinfo
  {journal} {Rev. Mod. Phys.}\ }\textbf {\bibinfo {volume} {62}} (\bibinfo
  {year} {1990}),\ 10.1103/RevModPhys.62.113}\BibitemShut {NoStop}%
\bibitem [{\citenamefont {Sablikov}(2017)}]{Sablikov}%
  \BibitemOpen
  \bibfield  {author} {\bibinfo {author} {\bibfnamefont {V.~A.}\ \bibnamefont
  {Sablikov}},\ }\href {\doibase 10.1103/PhysRevB.95.085417} {\bibfield
  {journal} {\bibinfo  {journal} {Phys. Rev. B}\ }\textbf {\bibinfo {volume}
  {95}},\ \bibinfo {pages} {085417} (\bibinfo {year} {2017})}\BibitemShut
  {NoStop}%
\bibitem [{\citenamefont {Downing}\ and\ \citenamefont
  {Portnoi}(2017)}]{Portnoi}%
  \BibitemOpen
  \bibfield  {author} {\bibinfo {author} {\bibfnamefont {C.~A.}\ \bibnamefont
  {Downing}}\ and\ \bibinfo {author} {\bibfnamefont {M.~E.}\ \bibnamefont
  {Portnoi}},\ }\href {\doibase 10.1038/s41467-017-00949-y} {\bibfield
  {journal} {\bibinfo  {journal} {Nature Communications}\ }\textbf {\bibinfo
  {volume} {8}},\ \bibinfo {pages} {897} (\bibinfo {year} {2017})}\BibitemShut
  {NoStop}%
\bibitem [{\citenamefont {Sabio}\ \emph {et~al.}(2010)\citenamefont {Sabio},
  \citenamefont {Sols},\ and\ \citenamefont {Guinea}}]{Sabio2010}%
  \BibitemOpen
  \bibfield  {author} {\bibinfo {author} {\bibfnamefont {J.}~\bibnamefont
  {Sabio}}, \bibinfo {author} {\bibfnamefont {F.}~\bibnamefont {Sols}}, \ and\
  \bibinfo {author} {\bibfnamefont {F.}~\bibnamefont {Guinea}},\ }\href
  {\doibase 10.1103/PhysRevB.81.045428} {\bibfield  {journal} {\bibinfo
  {journal} {Phys. Rev. B}\ }\textbf {\bibinfo {volume} {81}},\ \bibinfo
  {pages} {045428} (\bibinfo {year} {2010})}\BibitemShut {NoStop}%
\bibitem [{\citenamefont {Lee}\ \emph {et~al.}(2012)\citenamefont {Lee},
  \citenamefont {Milstein},\ and\ \citenamefont {Terekhov}}]{Lee2012}%
  \BibitemOpen
  \bibfield  {author} {\bibinfo {author} {\bibfnamefont {R.~N.}\ \bibnamefont
  {Lee}}, \bibinfo {author} {\bibfnamefont {A.~I.}\ \bibnamefont {Milstein}}, \
  and\ \bibinfo {author} {\bibfnamefont {I.~S.}\ \bibnamefont {Terekhov}},\
  }\href {\doibase 10.1103/PhysRevB.86.035425} {\bibfield  {journal} {\bibinfo
  {journal} {Phys. Rev. B}\ }\textbf {\bibinfo {volume} {86}},\ \bibinfo
  {pages} {035425} (\bibinfo {year} {2012})}\BibitemShut {NoStop}%
\bibitem [{\citenamefont {Mahmoodian}\ and\ \citenamefont
  {Entin}(2013)}]{MahmoodianEntinEPL2013}%
  \BibitemOpen
  \bibfield  {author} {\bibinfo {author} {\bibfnamefont {M.~M.}\ \bibnamefont
  {Mahmoodian}}\ and\ \bibinfo {author} {\bibfnamefont {M.~V.}\ \bibnamefont
  {Entin}},\ }\href {http://stacks.iop.org/0295-5075/102/i=3/a=37012}
  {\bibfield  {journal} {\bibinfo  {journal} {EPL (Europhysics Letters)}\
  }\textbf {\bibinfo {volume} {102}},\ \bibinfo {pages} {37012} (\bibinfo
  {year} {2013})}\BibitemShut {NoStop}%
\bibitem [{\citenamefont {Marnham}\ and\ \citenamefont
  {Shytov}(2015)}]{MarnhamShytovPRB2015}%
  \BibitemOpen
  \bibfield  {author} {\bibinfo {author} {\bibfnamefont {L.~L.}\ \bibnamefont
  {Marnham}}\ and\ \bibinfo {author} {\bibfnamefont {A.~V.}\ \bibnamefont
  {Shytov}},\ }\href {\doibase 10.1103/PhysRevB.92.085409} {\bibfield
  {journal} {\bibinfo  {journal} {Phys. Rev. B}\ }\textbf {\bibinfo {volume}
  {92}},\ \bibinfo {pages} {085409} (\bibinfo {year} {2015})}\BibitemShut
  {NoStop}%
\bibitem [{\citenamefont {Hartmann}\ \emph {et~al.}(2011)\citenamefont
  {Hartmann}, \citenamefont {Shelykh},\ and\ \citenamefont
  {Portnoi}}]{Hartmann}%
  \BibitemOpen
  \bibfield  {author} {\bibinfo {author} {\bibfnamefont {R.~R.}\ \bibnamefont
  {Hartmann}}, \bibinfo {author} {\bibfnamefont {I.~A.}\ \bibnamefont
  {Shelykh}}, \ and\ \bibinfo {author} {\bibfnamefont {M.~E.}\ \bibnamefont
  {Portnoi}},\ }\href {\doibase 10.1103/PhysRevB.84.035437} {\bibfield
  {journal} {\bibinfo  {journal} {Phys. Rev. B}\ }\textbf {\bibinfo {volume}
  {84}},\ \bibinfo {pages} {035437} (\bibinfo {year} {2011})}\BibitemShut
  {NoStop}%
\bibitem [{\citenamefont {Gindikin}\ and\ \citenamefont
  {Sablikov}(2018)}]{PhysRevB.98.115137}%
  \BibitemOpen
  \bibfield  {author} {\bibinfo {author} {\bibfnamefont {Y.}~\bibnamefont
  {Gindikin}}\ and\ \bibinfo {author} {\bibfnamefont {V.~A.}\ \bibnamefont
  {Sablikov}},\ }\href {\doibase 10.1103/PhysRevB.98.115137} {\bibfield
  {journal} {\bibinfo  {journal} {Phys. Rev. B}\ }\textbf {\bibinfo {volume}
  {98}},\ \bibinfo {pages} {115137} (\bibinfo {year} {2018})}\BibitemShut
  {NoStop}%
\bibitem [{\citenamefont {Winkler}\ \emph {et~al.}(2006)\citenamefont
  {Winkler}, \citenamefont {Thalhammer}, \citenamefont {Lang}, \citenamefont
  {Grimm}, \citenamefont {Denschlag}, \citenamefont {Daley}, \citenamefont
  {Kantian}, \citenamefont {B{\"u}chler},\ and\ \citenamefont
  {Zoller}}]{winkler2006repulsively}%
  \BibitemOpen
  \bibfield  {author} {\bibinfo {author} {\bibfnamefont {K.}~\bibnamefont
  {Winkler}}, \bibinfo {author} {\bibfnamefont {G.}~\bibnamefont {Thalhammer}},
  \bibinfo {author} {\bibfnamefont {F.}~\bibnamefont {Lang}}, \bibinfo {author}
  {\bibfnamefont {R.}~\bibnamefont {Grimm}}, \bibinfo {author} {\bibfnamefont
  {J.~H.}\ \bibnamefont {Denschlag}}, \bibinfo {author} {\bibfnamefont {A.~J.}\
  \bibnamefont {Daley}}, \bibinfo {author} {\bibfnamefont {A.}~\bibnamefont
  {Kantian}}, \bibinfo {author} {\bibfnamefont {H.~P.}\ \bibnamefont
  {B{\"u}chler}}, \ and\ \bibinfo {author} {\bibfnamefont {P.}~\bibnamefont
  {Zoller}},\ }\href {\doibase 10.1038/nature04918} {\bibfield  {journal}
  {\bibinfo  {journal} {Nature}\ }\textbf {\bibinfo {volume} {441}},\ \bibinfo
  {pages} {853} (\bibinfo {year} {2006})}\BibitemShut {NoStop}%
\bibitem [{\citenamefont {Zhou}\ \emph {et~al.}(2014)\citenamefont {Zhou},
  \citenamefont {Wang},\ and\ \citenamefont {Wang}}]{PhysRevB.89.195119}%
  \BibitemOpen
  \bibfield  {author} {\bibinfo {author} {\bibfnamefont {S.}~\bibnamefont
  {Zhou}}, \bibinfo {author} {\bibfnamefont {Y.}~\bibnamefont {Wang}}, \ and\
  \bibinfo {author} {\bibfnamefont {Z.}~\bibnamefont {Wang}},\ }\href {\doibase
  10.1103/PhysRevB.89.195119} {\bibfield  {journal} {\bibinfo  {journal} {Phys.
  Rev. B}\ }\textbf {\bibinfo {volume} {89}},\ \bibinfo {pages} {195119}
  (\bibinfo {year} {2014})}\BibitemShut {NoStop}%
\bibitem [{\citenamefont {Han}\ \emph {et~al.}(2016)\citenamefont {Han},
  \citenamefont {Liu}, \citenamefont {Liu}, \citenamefont {Li}, \citenamefont
  {Chen}, \citenamefont {Liao}, \citenamefont {Xie}, \citenamefont {Normand},\
  and\ \citenamefont {Xiang}}]{Han_2016}%
  \BibitemOpen
  \bibfield  {author} {\bibinfo {author} {\bibfnamefont {X.-J.}\ \bibnamefont
  {Han}}, \bibinfo {author} {\bibfnamefont {Y.}~\bibnamefont {Liu}}, \bibinfo
  {author} {\bibfnamefont {Z.-Y.}\ \bibnamefont {Liu}}, \bibinfo {author}
  {\bibfnamefont {X.}~\bibnamefont {Li}}, \bibinfo {author} {\bibfnamefont
  {J.}~\bibnamefont {Chen}}, \bibinfo {author} {\bibfnamefont {H.-J.}\
  \bibnamefont {Liao}}, \bibinfo {author} {\bibfnamefont {Z.-Y.}\ \bibnamefont
  {Xie}}, \bibinfo {author} {\bibfnamefont {B.}~\bibnamefont {Normand}}, \ and\
  \bibinfo {author} {\bibfnamefont {T.}~\bibnamefont {Xiang}},\ }\href
  {\doibase 10.1088/1367-2630/18/10/103004} {\bibfield  {journal} {\bibinfo
  {journal} {New Journal of Physics}\ }\textbf {\bibinfo {volume} {18}},\
  \bibinfo {pages} {103004} (\bibinfo {year} {2016})}\BibitemShut {NoStop}%
\bibitem [{\citenamefont {{Gross}}\ \emph {et~al.}(1971)\citenamefont
  {{Gross}}, \citenamefont {{Perel'}},\ and\ \citenamefont
  {{Shekhmamet'ev}}}]{Gross}%
  \BibitemOpen
  \bibfield  {author} {\bibinfo {author} {\bibfnamefont {E.~F.}\ \bibnamefont
  {{Gross}}}, \bibinfo {author} {\bibfnamefont {V.~I.}\ \bibnamefont
  {{Perel'}}}, \ and\ \bibinfo {author} {\bibfnamefont {R.~I.}\ \bibnamefont
  {{Shekhmamet'ev}}},\ }\href@noop {} {\bibfield  {journal} {\bibinfo
  {journal} {Soviet Journal of Experimental and Theoretical Physics Letters}\
  }\textbf {\bibinfo {volume} {13}},\ \bibinfo {pages} {229} (\bibinfo {year}
  {1971})}\BibitemShut {NoStop}%
\bibitem [{\citenamefont {Sablikov}\ and\ \citenamefont
  {Shchamkhalova}(2019)}]{Sablikov1}%
  \BibitemOpen
  \bibfield  {author} {\bibinfo {author} {\bibfnamefont {V.~A.}\ \bibnamefont
  {Sablikov}}\ and\ \bibinfo {author} {\bibfnamefont {B.~S.}\ \bibnamefont
  {Shchamkhalova}},\ }\href {\doibase 10.1002/pssr.201900358} {\bibfield
  {journal} {\bibinfo  {journal} {physica status solidi (RRL) – Rapid
  Research Letters}\ }\textbf {\bibinfo {volume} {13}},\ \bibinfo {pages}
  {1900358} (\bibinfo {year} {2019})}\BibitemShut {NoStop}%
\bibitem [{\citenamefont {Hanamura}(1988)}]{Hanamura}%
  \BibitemOpen
  \bibfield  {author} {\bibinfo {author} {\bibfnamefont {E.}~\bibnamefont
  {Hanamura}},\ }\href {\doibase 10.1103/PhysRevB.38.1228} {\bibfield
  {journal} {\bibinfo  {journal} {Phys. Rev. B}\ }\textbf {\bibinfo {volume}
  {38}},\ \bibinfo {pages} {1228} (\bibinfo {year} {1988})}\BibitemShut
  {NoStop}%
\bibitem [{\citenamefont {Andreani}\ \emph {et~al.}(1991)\citenamefont
  {Andreani}, \citenamefont {Tassone},\ and\ \citenamefont
  {Bassani}}]{Andreani}%
  \BibitemOpen
  \bibfield  {author} {\bibinfo {author} {\bibfnamefont {L.}~\bibnamefont
  {Andreani}}, \bibinfo {author} {\bibfnamefont {F.}~\bibnamefont {Tassone}}, \
  and\ \bibinfo {author} {\bibfnamefont {F.}~\bibnamefont {Bassani}},\ }\href
  {\doibase https://doi.org/10.1016/0038-1098(91)90761-J} {\bibfield  {journal}
  {\bibinfo  {journal} {Solid State Communications}\ }\textbf {\bibinfo
  {volume} {77}},\ \bibinfo {pages} {641 } (\bibinfo {year}
  {1991})}\BibitemShut {NoStop}%
\bibitem [{\citenamefont {Robert}\ \emph {et~al.}(2016)\citenamefont {Robert},
  \citenamefont {Lagarde}, \citenamefont {Cadiz}, \citenamefont {Wang},
  \citenamefont {Lassagne}, \citenamefont {Amand}, \citenamefont {Balocchi},
  \citenamefont {Renucci}, \citenamefont {Tongay}, \citenamefont {Urbaszek},\
  and\ \citenamefont {Marie}}]{Robert}%
  \BibitemOpen
  \bibfield  {author} {\bibinfo {author} {\bibfnamefont {C.}~\bibnamefont
  {Robert}}, \bibinfo {author} {\bibfnamefont {D.}~\bibnamefont {Lagarde}},
  \bibinfo {author} {\bibfnamefont {F.}~\bibnamefont {Cadiz}}, \bibinfo
  {author} {\bibfnamefont {G.}~\bibnamefont {Wang}}, \bibinfo {author}
  {\bibfnamefont {B.}~\bibnamefont {Lassagne}}, \bibinfo {author}
  {\bibfnamefont {T.}~\bibnamefont {Amand}}, \bibinfo {author} {\bibfnamefont
  {A.}~\bibnamefont {Balocchi}}, \bibinfo {author} {\bibfnamefont
  {P.}~\bibnamefont {Renucci}}, \bibinfo {author} {\bibfnamefont
  {S.}~\bibnamefont {Tongay}}, \bibinfo {author} {\bibfnamefont
  {B.}~\bibnamefont {Urbaszek}}, \ and\ \bibinfo {author} {\bibfnamefont
  {X.}~\bibnamefont {Marie}},\ }\href {\doibase 10.1103/PhysRevB.93.205423}
  {\bibfield  {journal} {\bibinfo  {journal} {Phys. Rev. B}\ }\textbf {\bibinfo
  {volume} {93}},\ \bibinfo {pages} {205423} (\bibinfo {year}
  {2016})}\BibitemShut {NoStop}%
\bibitem [{\citenamefont {Wang}\ \emph {et~al.}(2018)\citenamefont {Wang},
  \citenamefont {Chernikov}, \citenamefont {Glazov}, \citenamefont {Heinz},
  \citenamefont {Marie}, \citenamefont {Amand},\ and\ \citenamefont
  {Urbaszek}}]{Glazov}%
  \BibitemOpen
  \bibfield  {author} {\bibinfo {author} {\bibfnamefont {G.}~\bibnamefont
  {Wang}}, \bibinfo {author} {\bibfnamefont {A.}~\bibnamefont {Chernikov}},
  \bibinfo {author} {\bibfnamefont {M.~M.}\ \bibnamefont {Glazov}}, \bibinfo
  {author} {\bibfnamefont {T.~F.}\ \bibnamefont {Heinz}}, \bibinfo {author}
  {\bibfnamefont {X.}~\bibnamefont {Marie}}, \bibinfo {author} {\bibfnamefont
  {T.}~\bibnamefont {Amand}}, \ and\ \bibinfo {author} {\bibfnamefont
  {B.}~\bibnamefont {Urbaszek}},\ }\href {\doibase
  10.1103/RevModPhys.90.021001} {\bibfield  {journal} {\bibinfo  {journal}
  {Rev. Mod. Phys.}\ }\textbf {\bibinfo {volume} {90}},\ \bibinfo {pages}
  {021001} (\bibinfo {year} {2018})}\BibitemShut {NoStop}%
\bibitem [{\citenamefont {Bernevig}\ \emph {et~al.}(2006)\citenamefont
  {Bernevig}, \citenamefont {Hughes},\ and\ \citenamefont {Zhang}}]{BHZ}%
  \BibitemOpen
  \bibfield  {author} {\bibinfo {author} {\bibfnamefont {B.~A.}\ \bibnamefont
  {Bernevig}}, \bibinfo {author} {\bibfnamefont {T.~L.}\ \bibnamefont
  {Hughes}}, \ and\ \bibinfo {author} {\bibfnamefont {S.-C.}\ \bibnamefont
  {Zhang}},\ }\href {\doibase 10.1126/science.1133734} {\bibfield  {journal}
  {\bibinfo  {journal} {Science}\ }\textbf {\bibinfo {volume} {314}},\ \bibinfo
  {pages} {1757} (\bibinfo {year} {2006})}\BibitemShut {NoStop}%
\bibitem [{\citenamefont {Glazov}\ and\ \citenamefont {Suris}(2020)}]{Suris}%
  \BibitemOpen
  \bibfield  {author} {\bibinfo {author} {\bibfnamefont {M.}~\bibnamefont
  {Glazov}}\ and\ \bibinfo {author} {\bibfnamefont {R.}~\bibnamefont {Suris}},\
  }\href {\doibase 10.3367/UFNe.2019.10.038663} {\bibfield  {journal} {\bibinfo
   {journal} {Phys. Usp.}\ }\textbf {\bibinfo {volume} {accepted}},\ \bibinfo
  {pages} {1} (\bibinfo {year} {2020})}\BibitemShut {NoStop}%
\end{thebibliography}%
\end{document}